\documentclass[aps,prl,twocolumn,superscriptaddress,longbibliography,nofootinbib,10pt]{revtex4-2}

\usepackage{amsmath,amssymb,graphicx}
\usepackage[colorlinks=true,linkcolor=blue,citecolor=blue]{hyperref}

\providecommand{\ket}[1]{\left|#1\right\rangle}
\providecommand{\bra}[1]{\left\langle#1\right|}
\newcommand{\Nn}{\mathcal{N}}
\newcommand{\ee}{\mathrm{e}}

\newcounter{smsec}
\renewcommand{\thesmsec}{S\arabic{smsec}}
\newcommand{\smsection}[1]{%
  \refstepcounter{smsec}\section*{\thesmsec.\quad #1}}

\begin{document}

\title{Exact solution of the forward-scattering approximation for the PXP model}

\author{Saurav Pandey}
\author{Yevgeny Bar Lev}
\affiliation{Department of Physics, Ben-Gurion University of the Negev,
Beer-Sheva 8410501, Israel}

\date{\today}

\begin{abstract}
The forward-scattering approximation (FSA) for the PXP scar tower is a
tight-binding chain whose hopping coefficients were known only
numerically. We show that they follow exactly from a one-dimensional
hard-rod gas, with weights set by alternating permutations that count
blockade-allowed spin-flip orderings. The equation of state of the gas
gives a limiting hopping profile that differs from the spin-$L/2$ form,
and the spectrum remains globally anharmonic. Yet within the FSA, the revivals of the N\'eel state become perfect as $L\to\infty$, and their number above any fixed fidelity grows as $\sqrt{L}$.
\end{abstract}

\maketitle


Generic isolated quantum many-body systems are expected to thermalize under their own unitary dynamics. This expectation is formalized by the eigenstate thermalization hypothesis (ETH)~\cite{Deutsch1991,Srednicki1994,Rigol2008,DAlessio2016}, which states that individual eigenstates in the bulk of the spectrum already appear thermal to local probes. Integrable and many-body localized systems~\cite{Nandkishore2015,Abanin2019} are the standard exceptions to ETH and violate it for essentially all eigenstates. A weaker violation occurs when a nonintegrable system hosts a small number of atypical eigenstates embedded in an otherwise thermal spectrum~\cite{Shiraishi2017,Moudgalya2018}. Such states are called quantum many-body scars, and their presence is referred to as weak ergodicity breaking~\cite{Turner2018a,Serbyn2021,Moudgalya2022,Chandran2023}. Since the scars make up a vanishing fraction of the spectrum, typical initial states still thermalize, whereas states with substantial overlap on the scars show long-lived oscillations.

The canonical example of weak ergodicity breaking is a chain of Rydberg
atoms. In the experiment of Bernien \emph{et al.}~\cite{Bernien2017}, 51
atoms prepared in the alternating N\'eel configuration showed long-lived
oscillations of local observables. These oscillations were subsequently
attributed to a tower of nearly equally spaced scar eigenstates of the PXP
model~\cite{Turner2018a,Turner2018b}, which describes the atoms in the
limit of perfect Rydberg blockade~\cite{Fendley2004,Lesanovsky2012}. The
PXP model is nonintegrable~\cite{Park2025}, and in contrast to models
constructed to host exact scars~\cite{Shiraishi2017,Moudgalya2018}, its
scars are not protected by any known symmetry. Their origin has been
discussed in terms of an approximate $\mathrm{su}(2)$
algebra~\cite{Choi2019} and periodic orbits in a variational
manifold~\cite{Ho2019,Michailidis2020}.

The standard analytic handle on the PXP scar tower is the
forward-scattering approximation (FSA)~\cite{Turner2018a,Turner2018b},
which projects the PXP Hamiltonian of a chain of $L$ atoms onto an
approximately invariant subspace of dimension $L+1$. This yields a
tight-binding chain with hopping coefficients $\beta_n$, whose eigenvalues
and eigenvectors approximate the scar tower well at accessible system
sizes~\cite{Turner2018b}. Computing the $\beta_n$, however, requires building the basis states of the FSA subspace in a constrained Hilbert space whose dimension grows exponentially with $L$~\cite{Turner2018b}. Even within the FSA, it thus remains open whether the spectrum becomes harmonic as $L\to\infty$ and what happens to the N\'eel-state revivals in that limit.

In this Letter, we show that the FSA hopping coefficients follow exactly
from a one-dimensional gas of hard rods. The $\beta_n$ are fixed by the
norms of the FSA basis states, which count the blockade-allowed spin-flip
orderings from the N\'eel state. The fraction of allowed orderings
factorizes over domains of flipped sites, each weighted by the number of
alternating permutations of its length~\cite{Andre1879,Stanley2010}. 
The domains behave as hard rods, which makes the FSA chain exactly
computable at any $L$ and its thermodynamic limit analytically
accessible. The limiting hopping profile is fixed by the equation of state
of the gas, and the FSA spectrum remains globally anharmonic as
$L\to\infty$. The revivals nevertheless survive: each FSA revival becomes perfect in the
thermodynamic limit, and their number above any fixed fidelity threshold grows as
$\sqrt L$.

\emph{The FSA chain.}---We consider an open chain of $L$ sites, with $L$ even. Each site $i$ hosts an atom in either its ground state $\ket{0}_i$ or its Rydberg state $\ket{1}_i$, which we refer to as empty and occupied, respectively. The PXP Hamiltonian reads~\cite{Bernien2017,Turner2018a}
\begin{equation}
H=\sum_{i=1}^{L}P_{i-1}X_iP_{i+1},
\label{eq:pxp}
\end{equation}
where $X_i=\ket{0}_i\bra{1}_i+\ket{1}_i\bra{0}_i$ flips site $i$ and $P_i=\ket{0}_i\bra{0}_i$ projects it onto the empty state, with $P_0=P_{L+1}=1$ at the boundaries. The term $P_{i-1}X_iP_{i+1}$ flips site $i$ only if both of its neighbors are empty, and otherwise annihilates the state. Starting from the N\'eel state, $\ket{v_0} = \ket{1010\cdots10}$, the dynamics is therefore confined to a constrained Hilbert space, spanned by configurations with no two adjacent occupied sites, whose dimension is the Fibonacci number $F_{L+2}\sim\phi^{L}$, with $\phi=(1+\sqrt5)/2$~\cite{Turner2018b}.

Each local term of the Hamiltonian in Eq.~\eqref{eq:pxp} flips exactly one site of a configuration and therefore changes its Hamming distance from $\ket{v_0}$ by $\pm1$. We may thus split the Hamiltonian as $H=H^{+}+H^{-}$, with $H^{-}=(H^{+})^{\dagger}$, where $H^{+}$ collects the moves that increase the Hamming distance from $\ket{v_0}$. It is useful to separate the chain into two sublattices, $A$ of odd sites and $B$ of even sites, which are respectively occupied and empty in $\ket{v_0}$. Under $H^{+}$, a site on the $A$ sublattice can only flip from $1$ to $0$, while a site on the $B$ sublattice can only flip from $0$ to $1$. Consequently, in a product $(H^{+})^{n}$, no site can flip twice, since the second flip of any site would return it to its value in $\ket{v_0}$, thereby decreasing the Hamming distance. The vector $(H^{+})^{n}\ket{v_0}$ is therefore a superposition of configurations at Hamming distance $n$ from $\ket{v_0}$, and the normalized vectors
\begin{equation}
\ket{v_n}=\frac{(H^{+})^{n}\ket{v_0}}{\sqrt{\Nn_n}},
\qquad
\Nn_n\equiv\big\|(H^{+})^{n}\ket{v_0}\big\|^{2},
\label{eq:vn}
\end{equation}
which live in distinct Hamming shells, are orthogonal by construction. They form the FSA basis for $n=0,\dots,L$, and the ladder terminates at $n=L$, where a single configuration is left, $\ket{v_L}=\ket{0101\cdots01}$, the exact complement of $\ket{v_0}$. Projecting $H$ onto the span of $\{\ket{v_n}\}$ gives a bare hopping chain on $L+1$ sites,
\begin{equation}
H_{\rm FSA}=\sum_{n=0}^{L-1}\beta_n\Big(\ket{v_{n+1}}\bra{v_n}+\text{h.c.}\Big),
\;
\beta_n=\sqrt{\frac{\Nn_{n+1}}{\Nn_n}}.
\label{eq:fsachain}
\end{equation}
The FSA chain is thus fixed entirely by the norms $\Nn_0,\dots,\Nn_L$.

\emph{Path counting.}---To compute the norms, let us write $H^{+}=\sum_i g_i$, with $g_i$ the term acting at site $i$,
so that $(H^{+})^{n}=\sum_{i_1,\dots,i_n}g_{i_n}\cdots g_{i_1}$, with each term labeled by a string $i_1,\dots,i_n$ of sites. Since $g_i$ flips site $i$ and acts diagonally on its neighbors through
the projectors, it either returns the flipped configuration with amplitude $+1$, or annihilates it if site $i$ has
already flipped away from its value in $\ket{v_0}$ or if the blockade forbids the move. Acting on $\ket{v_0}$, a string thus specifies the order in which the $n$ sites flip, and survives only if the sites are distinct and every
flip is blockade-allowed in that order. Every surviving string, which we call a \emph{path}, therefore contributes a configuration at Hamming distance $n$ with amplitude $+1$, and the total amplitude of a configuration in $(H^{+})^{n}\ket{v_0}$ is simply the number of paths reaching it from $\ket{v_0}$.

To make the argument precise, let $\ket{S}$ denote a configuration
obtained from $\ket{v_0}$ by flipping $n$ sites, and let
$S\subseteq\{1,\ldots,L\}$ be the set containing the indices of the flipped
sites. We call $S$ a valid set if $\ket{S}$ contains no two adjacent
occupied sites. Of the $n!$ orderings of the flips that take $\ket{v_0}$
to $\ket{S}$, let $\varphi(S)$ denote the fraction that are paths, that
is, the probability that an ordering drawn uniformly at random respects
the blockade. The amplitude of $\ket{S}$ is then $n!\,\varphi(S)$, so that
\begin{equation}
(H^{+})^{n}\ket{v_0}=n!\sum_{|S|=n}\varphi(S)\ket{S},
\label{eq:Hplusn}
\end{equation}
and, since the $\ket{S}$ are orthonormal, Eqs.~\eqref{eq:vn} and~\eqref{eq:fsachain} give
\begin{align}
\Nn_n&=(n!)^{2}Z_n,
\quad
Z_n=\sum_{|S|=n}\varphi(S)^{2},
\label{eq:NZ}\\
\beta_n&=(n+1)\,\sqrt{Z_{n+1}/Z_n}.
\label{eq:betaZ}
\end{align}
Computing $Z_n$ requires determining the fractions $\varphi(S)$ and then summing their squares over all valid sets $S$ of $n$ flipped sites.

\emph{Zigzag orderings.}---The fractions $\varphi(S)$ are determined by
how the blockade restricts the order of the spin flips. Let $i\in S$ be
one of the flipped sites and let $\tau_i$ denote its position in an
ordering of all the spin flips in $S$. Since a flip requires the neighboring
sites to be empty, a neighbor $j\sim i$ can constrain $\tau_i$ in one of
four ways,
\begin{equation}
\begin{array}{l|cc}
 & j\in S & j\notin S\\ \hline
i\in A\ (1\to0) & \tau_i<\tau_j & \text{no constraint}\\
i\in B\ (0\to1) & \tau_j<\tau_i & \text{invalid $S$}
\end{array}
\label{eq:ebfrule}
\end{equation}
The first column of Eq.~\eqref{eq:ebfrule} tells us that on every bond whose
two sites both flip, the $A$ site that empties must flip before the $B$ site that
fills. We call this the \emph{empty-before-fill} rule. The second column, which corresponds to an unflipped neighboring site, has two consequences. First, the unflipped sites do not constrain the order of the flips. This motivates us to define a \emph{domain} as a maximal run of adjacent flipped sites, so that distinct domains are separated by at least one unflipped site. Second, for $S$ to be a valid set, a flipped $B$ site must have its $A$ neighbors also among the flipped sites of $S$. Consequently, a domain begins and ends on $A$ sites and has odd length $m=2r-1$, with $r$ sites on $A$ and $r-1$ on $B$. We call it a \emph{bulk domain}. The only exception occurs if site $L \in S$, which is a $B$ site with a single neighbor to its left. A domain ending there has even length $m=2r$, with $r$ sites on each sublattice, and we call it an \emph{edge domain}. 

By Eq.~\eqref{eq:ebfrule}, the only ordering constraints are between adjacent flipped sites, so the blockade acts only within domains. For any valid set $S$ with $|S|=n$, consider an ordering of the flips drawn uniformly at
random from the $n!$ possible ones. Since the orders within distinct
domains are independent, the probability $\varphi(S)$ that such an
ordering is allowed factorizes across the domains of $S$, and together with Eq.~\eqref{eq:NZ}, this gives
\begin{equation}
\varphi(S)=\prod_{k}p_{m_k},
\qquad
Z_n=\sum_{|S|=n}\ \prod_k p_{m_k}^{2},
\label{eq:factorize}
\end{equation}
where $m_1,m_2,\dots$ are the lengths of the domains of $S$, and $p_m$ is
the probability that a random order of the $m$ flips of a single domain
is allowed. A configuration thus enters $Z_n$ only through the lengths of
its domains.

To find $p_m$, consider a domain of length $m$ occupying sites $a,a+1,\dots,a+m-1$,
with $a\in A$. The empty-before-fill rule applied to each bond gives
$\tau_a<\tau_{a+1}>\tau_{a+2}<\tau_{a+3}>\cdots$, so the flip order must
zigzag along the domain. These inequalities involve only the relative order of the domain's $m$ flips. Thus
$p_m=\mathcal A_m/m!$, where $\mathcal A_m$ is the number of orders of $m$
flips that zigzag in this way, that is, the number of alternating
permutations~\cite{Andre1879,Stanley2010},
\begin{equation}
\sum_{m\ge0}\frac{\mathcal A_m}{m!}x^{m}=\sec x+\tan x,
\label{eq:andre}
\end{equation}
which gives $\mathcal A_m=1,1,2,5,16,61,\dots$ and hence
$p_m=1,\tfrac12,\tfrac13,\tfrac{5}{24},\tfrac{2}{15},\dots$ for $m\ge1$. Since $p_m>0$ for every $m$, every valid $S$ with $|S|=n$ contributes to $(H^{+})^{n}\ket{v_0}$, and the $p_m$ have closed forms~\cite{Flajolet2009} given in Sec.~\ref{sm:paths} of the Supplemental Material (SM)~\cite{supp}. What remains is to perform the sum in Eq.~\eqref{eq:factorize} over all valid sets at fixed $n$.

\begin{figure*}[t]
\centering
\includegraphics[width=\textwidth]{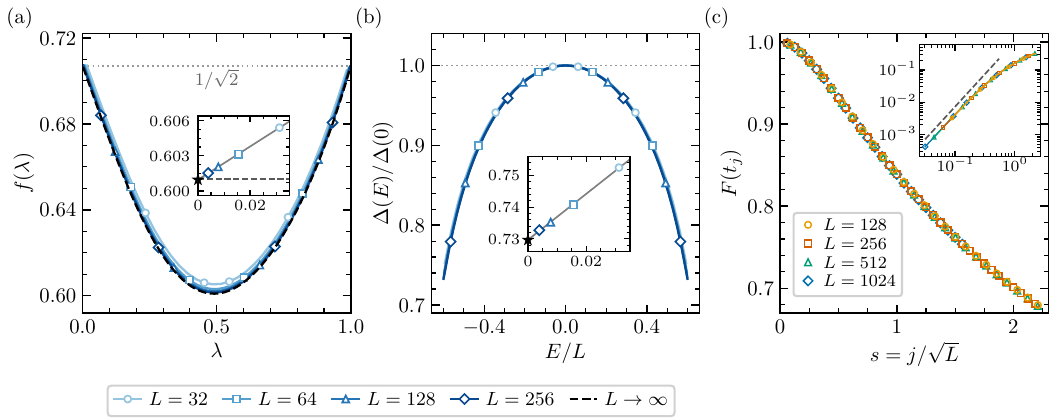}
\caption{(a) Hopping profile. Main panel: $f_n$ of Eq.~\eqref{eq:fdef},
obtained from the finite-$L$ recursion of Eqs.~\eqref{eq:grec}
and~\eqref{eq:transfer} and plotted at the bond midpoint
$\lambda=(n+\tfrac12)/L$, together with its thermodynamic limit,
Eq.~\eqref{eq:fprofile} (dashed). The dotted line marks the exact endpoint
value $f(0)=f(1)=1/\sqrt2$. Inset: $f(1/2)$ against $1/L$, extrapolated
linearly (star) and compared with the equation-of-state value (dashed).
(b) Level spacings. Main panel: $\Delta(E)/\Delta(0)$ of the FSA chain,
Eq.~\eqref{eq:fsachain}, against $E/L$. The dotted line marks uniform
spacing, $\Delta(E)=\Delta(0)$. Inset: the ratio $\Delta_{\min}/\Delta(0)$
of the smallest spacing, at the band edge, to the largest, at the center,
against $1/L$, extrapolated linearly (star). In (a) and (b), darker curves
correspond to larger $L$. (c) Fidelity revivals. Main panel: the peak values $F(t_j)$ of 
the N\'eel-state fidelity against the scaled revival index $s=j/\sqrt L$. Inset: the deficit $1-F(t_j)$ against $s$ on 
logarithmic axes, with the dashed line $\propto s^2$.}
\label{fig:main}
\end{figure*}

\emph{Hard rods.}---A bulk domain of length $2r-1$ ends on $A$, so the $B$ site to its right must remain unflipped and cannot be part of another domain. Together they form a \emph{footprint} of $2r$ sites, as does an
edge domain of length $2r$ by itself. Grouping the chain into $M=L/2$ cells, with cell $c$ consisting of sites $2c-1\in A$ and $2c\in B$, each footprint covers $r$ cells and is called a \emph{rod} of length $r$, while
the uncovered cells are called \emph{vacancies} (Fig.~\ref{fig:cells} in the End Matter). A valid set $S$ is therefore nothing but a way of laying down non-overlapping rods along a row of $M$ cells, which is the standard picture of a one-dimensional hard-rod lattice gas~\cite{Tonks1936,LiebMattis1966}. Since the FSA chain is indexed by the number $n$ of flipped sites, we take the particles of the gas to be the flipped sites rather than the rods
(SM, Sec.~\ref{sm:rods}). A bulk rod of length $r$ then carries $2r-1$ particles and has weight $w_r=p_{2r-1}^{2}$; an edge rod, which can only occupy the last $r$ cells, carries $2r$ particles and has weight $\tilde w_r=p_{2r}^{2}$; and a vacancy carries no particles and has weight $1$. The sum in
Eq.~\eqref{eq:factorize} thus becomes the canonical partition function of the gas,
\begin{equation}
Z_n=\sum_{N(\eta)=n}\;\prod_{\text{rods}\in\eta} w_{\rm rod},
\label{eq:Zgas}
\end{equation}
where $\eta$ is a configuration of rods and vacancies on the $M$ cells, $N(\eta)$ is the number of particles in $\eta$, and $w_{\rm rod}$ is $w_r$ or $\tilde w_r$ according to the rod species.

\emph{Finite-$L$ recursion.}---Every configuration in Eq.~\eqref{eq:Zgas}
is a sequence of rods and vacancies laid down from left to right. Let
$g_c[k]$ denote the total weight of the configurations built from bulk
rods and vacancies that fill the first $c$ cells with $k$ particles. Any
such configuration ends either in a vacancy, whose removal leaves $c-1$
cells and all $k$ particles at unchanged weight, or in a bulk rod of
length $r$, whose removal leaves $c-r$ cells and $k-2r+1$ particles and
reduces the weight by a factor $w_r$. Hence
\begin{equation}
g_c[k]=g_{c-1}[k]+\sum_{r\ge1}g_{c-r}[k-2r+1]\,w_r,
\label{eq:grec}
\end{equation}
with $g_0[k]=\delta_{k0}$ and $g_c[k]=0$ for $c<0$. For $c\ge1$, $g_c[k]$
vanishes unless $0\le k\le2c-1$, the upper bound being reached by a single
bulk rod filling all $c$ cells. A configuration on $M$ cells with $n$ particles either contains no edge
rod, in which case it is accounted for in $g_M[n]$, or contains one, of
some length $r$ and weight $\tilde w_r$, which then occupies the last $r$
cells. In the latter case, the rest of the configuration is one of bulk rods and vacancies
on the first $M-r$ cells with $n-2r$ particles, so
\begin{equation}
Z_n=g_M[n]+\sum_{r\ge1}g_{M-r}[n-2r]\,\tilde w_r.
\label{eq:transfer}
\end{equation}
A single sweep of Eq.~\eqref{eq:grec} delivers the whole sequence
$Z_0,\dots,Z_L$, and with it every $\beta_n$ from Eq.~\eqref{eq:betaZ}, in
$O(L^3)$ operations, in exact agreement with direct construction of the
FSA states (SM, Fig.~\ref{fig:check}).

\emph{Thermodynamic limit.}---We now turn to the
thermodynamic limit, where the gas can be solved analytically. A natural reference for the hopping profile is the spin-$L/2$ form $\beta_n^{2}=\Omega^{2}(n+1)(L-n)$ with constant $\Omega$, for which the spectrum is exactly
harmonic~\cite{Choi2019}. We therefore write
\begin{equation}
\beta_n^{2}=(n+1)(L-n)\,f_n^{2},
\label{eq:fdef}
\end{equation}
so that $f_n$ measures the deviation from the spin-$L/2$ form, and seek
its limit $f(\lambda)$ as a function of the position $\lambda=n/L\in[0,1]$
along the FSA chain. This amounts to extracting the large-$L$ behavior of
$Z_n$, for which we follow the standard treatment of one-dimensional
classical gases~\cite{LiebMattis1966}, applied here to rods of varying
length. With $x$ counting cells and $u$ the fugacity per particle, the
bulk rods alone determine the thermodynamic limit through
$W(x,u)=\sum_{r\ge1}w_r\,x^{r}u^{2r-1}$, which enters the equation of
state of the hard-rod gas (Appendix~\ref{app:eos})
\begin{equation}
x_*+W(x_*,u)=1,
\qquad
\rho(u)=-\frac{\partial\ln x_*(u)}{\partial\ln u},
\label{eq:eos}
\end{equation}
where $x_*(u)$ is the smallest positive solution of $x+W(x,u)=1$, and
$\rho$ is the mean density, in particles per cell, at fugacity $u$. A position $\lambda$ along the FSA chain corresponds to $n=\lambda L$ particles on $M=L/2$ cells, that is, to the canonical density $2\lambda$.
We define the fugacity $u_*(\lambda)$ through $\rho(u_*)=2\lambda$. By
equivalence of ensembles, the gas at this fugacity is dominated by
configurations of density $2\lambda$, so that $Z_{n+1}/Z_n\to1/u_*(\lambda)$
(Appendix~\ref{app:eos}).
Equations~\eqref{eq:betaZ} and~\eqref{eq:fdef} then give the limiting
hopping profile
\begin{equation}
f^{2}(\lambda)=\frac{\lambda}{(1-\lambda)\,u_*(\lambda)}.
\label{eq:fprofile}
\end{equation}

Figure~\ref{fig:main}(a) shows the limiting profile together with the
finite-$L$ curves for $f_n$. The profile runs from the exact value $1/\sqrt2$ at both ends (SM, Sec.~\ref{sm:ends}) down to a minimum $f_{\min}=0.600954$ at $\lambda=0.48989$, and is therefore weakly asymmetric under $\lambda\to1-\lambda$. The finite-$L$ curves converge to the limiting profile as $L$ increases, which we quantify at $\lambda=1/2$, where their linear extrapolation in $1/L$ gives $f(1/2)=0.601$, against $0.601001$ from Eq.~\eqref{eq:fprofile}, an independent check of the thermodynamic limit. Since $f(\lambda)$ is clearly nonconstant, the deviation from the spin-$L/2$ form is not a finite-size effect.

\emph{Anharmonicity and revivals.}---A nonconstant $f(\lambda)$ does not
by itself imply an anharmonic spectrum, since hopping chains with very different profiles can share the same, even equally spaced, 
spectrum~\cite{Vinet2012PST}. We therefore examine the level spacings directly. Diagonalizing the FSA chain with the
$\beta_n$ of the hard-rod mapping gives the eigenvalues $E_j$, and we
consider the gap $\Delta(E)=E_{j+1}-E_j$, at $E=(E_j+E_{j+1})/2$.
Figure~\ref{fig:main}(b) shows that the spacings are largest at the
band center and decrease towards the band edges, converging onto a nonuniform profile. The ratio $\Delta_{\min}/\Delta(0)$ of the smallest spacing, at the band edge, to the largest, at the center, extrapolates to $0.73$, well below unity, so the FSA spectrum remains globally anharmonic in the thermodynamic limit.

The return probability, or fidelity, of the N\'eel state within the FSA
is $F(t)=|\bra{v_0}\ee^{-iH_{\rm FSA}t}\ket{v_0}|^{2}$. A nonuniform
spectrum causes dephasing, and thus one would expect the fidelity revivals to decay in the thermodynamic limit. However, as shown in Appendix~\ref{app:ldos}, the N\'eel state has zero mean energy and energy variance $L/2$, so it lies at the center of the spectrum and occupies only $O(\sqrt L)$ of the $L+1$ eigenstates. Over that central
window, the relative variation of the level spacings vanishes as $1/L$. Compared with a
uniform ladder of spacing $\Delta(0)$, the relative phases of the occupied
eigenstates then spread by only $O(j/\sqrt L)$ after $j$ revival periods, so that significant dephasing is expected only when $j$ becomes of order $\sqrt L$, suggesting $s=j/\sqrt L$ as the natural
scaling variable.

In Fig.~\ref{fig:main}(c), the collapse of the fidelity
revival peaks for different system sizes confirms the scaling behavior
$F(t_j)=\mathcal{F}(s)$, and implies that the number of revivals above any
fixed fidelity threshold grows as $\sqrt L$. Its inset further shows that the
scaling function behaves as $1-\mathcal{F}\propto s^{2}$ for small $s$. At
fixed $j$, this gives a deficit of $O(j^2/L)$, so every individual revival
within the FSA becomes perfect as $L\to\infty$. The revival period in this limit follows from the limiting hopping profile, $T_\infty=4.81193$ (Appendix~\ref{app:period}).

\emph{Discussion.}---We have shown that the forward-scattering
approximation of the PXP model is exactly solvable. The mapping to the
hard-rod gas replaces the construction of the FSA states in an
exponentially large Hilbert space by a recursion that yields every
hopping coefficient in $O(L^3)$ operations. In the thermodynamic limit, the FSA spectrum remains globally anharmonic, yet the revivals of the N\'eel state become perfect, as the state samples only the asymptotically uniform center of the spectrum. Likewise, the N\'eel state is transferred to its complement $\ket{v_L}$ halfway between the revivals: the transfer peaks collapse onto the same scaling function $\mathcal F(s)$ and thus become perfect as $L\to\infty$ (SM, Sec.~\ref{sm:pst}). Perfect transfer between the ends of a hopping chain requires every eigenstate to have equal weight on the two ends, which, for a finite chain, is equivalent to mirror symmetry~\cite{Vinet2012PST,Kay2010}. The FSA chain, whose ends are $\ket{v_0}$ and $\ket{v_L}$, lacks this symmetry even as $L\to\infty$, since $f(\lambda)\neq f(1-\lambda)$. The FSA chain thus provides an example of asymptotically perfect transfer in a chain without mirror symmetry.

These results, however, hold only within the FSA. The FSA subspace is not exactly invariant under the PXP Hamiltonian, since the approximation fails on the return, $H^{-}\ket{v_{n+1}}=\beta_n\ket{v_n}+\ket{r_n}$, where $\ket{r_n}$ lies outside the span of $\{\ket{v_n}\}$ and measures the leakage out of the FSA subspace. This leakage is controlled by the commutator $[H^{+},H^{-}]$~\cite{Turner2018b}, and with the revivals and transfers becoming asymptotically perfect, it is the leakage that must ultimately decide their fate in the full PXP model. Whether the hard-rod picture also gives access to it is a natural question for future work. More broadly, it would be interesting to see if the path-counting picture developed in this work carries over to other kinetically constrained
models~\cite{Serbyn2021,Moudgalya2022}.

\emph{Acknowledgments.}---S.P. thanks Sunidhi Sen for useful discussions. Y.B. acknowledges support from the Israel Science Foundation (Grant No.~1304/23). S.P.\ used Claude Opus~5.5 (Anthropic) as a research and writing assistant and to develop and debug numerical code. All AI-assisted material was verified by S.P., and the authors take full responsibility for the content of this Letter.

\bibliography{refs}
\clearpage

\section*{End Matter}

\setcounter{secnumdepth}{1}
\appendix

\section{The hard-rod gas in the thermodynamic limit}
\label{app:eos}

\begin{figure}[t]
\centering
\includegraphics[width=\columnwidth]{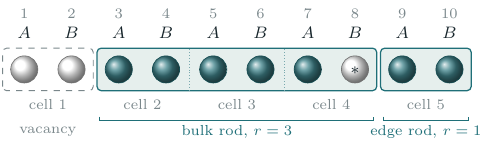}
\caption{The cell decomposition for $L=10$ and
$S=\{3,\dots,7\}\cup\{9,10\}$, that is, $\ket{S}=\ket{1001010001}$. Teal
spheres are flipped sites and white spheres are unflipped ones. The bulk
domain $\{3,\dots,7\}$ and the forced unflipped $B$ site $8$ (marked $*$)
fill cells $2$ to $4$ and constitute a bulk rod of length $r=3$. The edge
domain $\{9,10\}$ covers cell $5$ by itself and is thus an edge rod of
length $r=1$. Cell $1$, with no flipped site, is a vacancy.}
\label{fig:cells}
\end{figure}

In this appendix, we derive the equation of state of the hard-rod gas,
Eq.~\eqref{eq:eos}, and the limit $Z_{n+1}/Z_n\to1/u_*(\lambda)$ of the
ratio of the canonical partition function at consecutive particle numbers,
which leads to the limiting hopping profile of Eq.~\eqref{eq:fprofile}.
The gas is defined by the cell decomposition of the main text, which maps a 
valid set $S$ onto a configuration of rods and vacancies on $M=L/2$ cells, 
as illustrated in Fig.~\ref{fig:cells}.

To this end, we introduce two auxiliary variables $x$ and $u$, and collect every
$g_c[k]$ of Eq.~\eqref{eq:grec} into
$G(x,u)=\sum_{c\ge0}\sum_{k}g_c[k]\,x^{c}u^{k}$, so that $x$ counts cells
and $u$ counts particles. Multiplying Eq.~\eqref{eq:grec} by $x^{c}u^{k}$
and summing over both indices, the vacancy term contributes $xG$, the
factor $x$ appearing because the term is displaced by one cell. A bulk rod
of length $r$ is displaced by $r$ cells and $2r-1$ particles and thus
contributes $w_r x^{r}u^{2r-1}G$. Collecting the rod terms into
$W(x,u)=\sum_{r\ge1}w_r\,x^{r}u^{2r-1}$ and adding the initial condition
$g_0[k]=\delta_{k0}$, which contributes $1$, gives $G=1+xG+WG$, that is
\begin{equation}
G(x,u)=\frac{1}{1-x-W(x,u)}.
\label{eq:G}
\end{equation}

Equation~\eqref{eq:G} collects the weights of configurations made up of
bulk rods and vacancies only. It remains to account for configurations
with an edge rod. Since $Z_n$ is a canonical partition function at fixed
particle number, summing it over $n$ with a weight $u^{n}$ makes $u$ a
fugacity and gives the grand partition function of the gas on $M$ cells,
$\Xi_M(u)=\sum_{n}Z_nu^{n}$. Summing $\Xi_M$ over $M$ with a weight $x^{M}$ then collects every chain length
into a single object, and after substituting $Z_n$ from Eq.~\eqref{eq:transfer},
one obtains
\begin{equation}
\Xi(x,u)\equiv\sum_{M\ge0}\Xi_M(u)\,x^{M}=G(x,u)\,\tilde W(x,u),
\label{eq:Xi}
\end{equation}
where $\tilde W(x,u)=1+\sum_{r\ge1}\tilde w_r\,x^{r}u^{2r}$. Note that both
$W$ and $\tilde W$ involve $x$ and $u$ only through the combination $v=xu^{2}$, up to an overall factor of $u^{-1}$ in $W$,
so that
\begin{equation}
W(x,u)=u^{-1}Q(v),\qquad Q(v)=\sum_{r\ge1}w_r\,v^{r},
\label{eq:Qdef}
\end{equation}
and $\tilde W(x,u)=1+\sum_{r\ge1}\tilde w_r\,v^{r}$. The coefficients
fall off geometrically as $w_r\to4v_c^{-r}$ and
$\tilde w_r\to(4/\pi)^{2}v_c^{-r}$, with $v_c=(\pi/2)^{4}$ (SM,
Sec.~\ref{sm:paths}). Both series are therefore analytic for $v<v_c$ and
diverge at $v=v_c$.

With these generating functions, the mean density of the hard-rod gas can be computed from the grand
partition function in the usual way. At finite $M$,
$\partial\ln\Xi_M/\partial\ln u=\langle n\rangle_u$, so dividing by the
volume and taking the limit gives
$\rho(u)=\lim_{M\to\infty}M^{-1}\,\partial\ln\Xi_M/\partial\ln u$. Computing $\rho(u)$ thus
requires $\Xi_M(u)$ at large $M$. Since $\Xi(x,u)$ is the generating
function of $\Xi_M(u)$, the growth of $\Xi_M$ with $M$ is controlled by
the singularity of $\Xi$ in $x$ nearest to the origin. That singularity comes from the pole of $G$ in Eq.~\eqref{eq:G}, at the smallest positive root of $x+W(x,u)=1$, which, from Eq.~\eqref{eq:Qdef}, becomes $v+u\,Q(v)=u^{2}$. 
As the rod weights $w_r$ are positive, the left-hand side, $v+u\,Q(v)$, increases monotonically
from zero at $v=0$ and diverges as $v\to v_c$, where $Q$ diverges, so it
equals $u^{2}$ at a single $v_*<v_c$, which gives the root
$x_*(u)=v_*/u^{2}$. 
Note that the derivative $\partial_x(x+W)=1+u\,Q'(v) \geq1$ for all $v<v_c$,
which makes $x_*$ a simple pole of $G$, and the fact that $\tilde W$ is
regular for $v<v_c$ makes it the dominant singularity of $\Xi$.

A simple pole at $x_*$ gives $\Xi_M(u)\sim x_*(u)^{-M}$ up to a prefactor that is independent of $M$,
and substituting this into the definition of $\rho(u)$ yields the equation
of state, Eq.~\eqref{eq:eos}. 
Since $\tilde W$ is regular at $x_*$, it does not enter the growth rate of $\Xi_M$ with $M$,
which is why the edge rod weights drop out of the equation of state, and
only the bulk rods determine the thermodynamic limit.

It remains to connect the ratio $Z_{n+1}/Z_n$ to the fugacity,
in the limit $n,L\to\infty$ with $\lambda=n/L$ held fixed, which corresponds to the canonical
density $\rho=n/M=2\lambda$. At large $M$, the canonical partition
function of the gas is exponential in the volume, $Z_n\sim\ee^{Ms(\rho)}$,
which defines the entropy density $s(\rho)$. Stepping $n$ by one changes
$\rho$ by $1/M$, so that
\begin{equation}
\frac{Z_{n+1}}{Z_n}\ \rightarrow\
\exp\big(\partial s/\partial\rho\big)\big|_{\rho=2\lambda}.
\label{eq:dsdrho}
\end{equation}
To relate $\partial s/\partial\rho$ to the fugacity, we return to the
grand partition function. Substituting $Z_n\sim\ee^{Ms(\rho)}$ into
$\Xi_M(u)=\sum_nZ_nu^{n}$ and writing $u^{n}=\ee^{M\rho\ln u}$ gives
\begin{equation}
\Xi_M(u)\sim\sum_{\rho}\ee^{M\left[s(\rho)+\rho\ln u\right]}.
\label{eq:saddle}
\end{equation}
At large $M$, the sum in Eq.~\eqref{eq:saddle} is dominated by the density
$\rho$ that maximizes the exponent, which is given by the saddle point equation
$\partial s/\partial\rho+\ln u=0$. Since $x_*(u)$ is a simple root that
varies smoothly with $u$, the gas has no phase transition, and this
maximum is unique. By equivalence of ensembles, this maximizing density is
precisely the mean density $\rho(u)$ of Eq.~\eqref{eq:eos}. At the
fugacity $u=u_*(\lambda)$, where $\rho(u_*)=2\lambda$, the saddle point
equation becomes
$\partial s/\partial\rho|_{\rho=2\lambda}=-\ln u_*(\lambda)$, and
combining this with Eq.~\eqref{eq:dsdrho} yields
$Z_{n+1}/Z_n\to1/u_*(\lambda)$, as used in the main text.

\section{The N\'eel state in the FSA spectrum}
\label{app:ldos}

Here we establish the three results used in the discussion of the
revivals in the main text: the $O(\sqrt L)$ window of eigenstates occupied
by the N\'eel state, the $O(1/L)$ uniformity of the level spacings over
that window, and the resulting slow dephasing of the revivals.
We denote by $\langle\cdot\rangle_{v_0}$ the average over the FSA
eigenstates weighted by their overlaps $|\langle E_j|v_0\rangle|^{2}$ with
the N\'eel state, so that
$\langle h(E)\rangle_{v_0}=\bra{v_0}h(H_{\rm FSA})\ket{v_0}$. Since
$H_{\rm FSA}\ket{v_0}=\beta_0\ket{v_1}$ by Eq.~\eqref{eq:fsachain}, it
follows that $\langle E\rangle_{v_0}=0$ and that the energy variance is
\begin{equation}
\langle E^2\rangle_{v_0}=\beta_0^{2}=\Nn_1=\frac{L}{2},
\label{eq:sumrule}
\end{equation}
exactly at every $L$. The value $\Nn_1=L/2$ follows from the fact that
$H^{+}\ket{v_0}$ is a sum of the $L/2$ configurations obtained by emptying
a single $A$ site, each with amplitude $1$. As the energy bandwidth is
$O(L)$, the N\'eel state thus occupies only the central $O(\sqrt L)$ of the
$L+1$ eigenstates.

The spacings over that window can be estimated directly. The FSA chain has
no diagonal part, so it is chirally symmetric, and its spectrum is
invariant under $E\to-E$. Hence, $\Delta(E)$ is even, and since $\Delta(E)/\Delta(0)$ converges to a
function of $E/L$ [Fig.~\ref{fig:main}(b)], near the band center we may
expand it as
\begin{equation}
\frac{\Delta(E)}{\Delta(0)}=1-c_2\Big(\frac{E}{L}\Big)^{2}+\dots
\label{eq:quadratic}
\end{equation}
with no linear term, by the evenness, and with $c_2$ independent of $L$. Equation~\eqref{eq:sumrule} then gives 
\begin{equation}
\left\langle\frac{\Delta(E)}{\Delta(0)}\right\rangle_{v_0}\simeq1-\frac{c_2}{2L}.
\label{eq:flatwindow}
\end{equation}
Thus, although the spectrum remains globally anharmonic, the spacings
sampled by the N\'eel state become asymptotically uniform, with their
leading relative variation vanishing as $1/L$.

To see how this affects the fidelity revivals, we compare the FSA energy
window sampled by the N\'eel state with a uniform ladder of spacing
$\Delta(0)$. Accumulating the deviations of $O(\Delta(0)/L)$ in the level spacings
across the occupied window of $O(\sqrt L)$ width gives an energy mismatch
of $\delta E=O(\Delta(0)/\sqrt L)$. During time evolution, each FSA
eigenstate in the N\'eel-state superposition acquires a phase
$\ee^{-iE_jt}$. For a uniform ladder, these phases would realign perfectly
with revival period $T=2\pi/\Delta(0)$. The energy mismatch instead
produces a spread in their relative phases such that after $j$ periods,
\begin{equation}
\delta\phi_j\sim\delta E\times jT=O\Big(\frac{j}{\sqrt L}\Big),
\label{eq:deltaphi}
\end{equation}
which is the scaling used in the main text.

\section{The revival period}
\label{app:period}

The revival period in the thermodynamic limit can be determined
directly from the limiting hopping profile of Eq.~\eqref{eq:fprofile}. The N\'eel state forms a wavepacket of the FSA eigenstates drawn from the center of the band, where the local group velocity takes its maximal value $2\beta_n$, so semiclassically one may write $dn/dt=2\beta_n$. Rewriting this in terms of $\lambda=n/L$
using Eq.~\eqref{eq:fdef} gives
$d\lambda/dt=2\beta_n/L=2\sqrt{\lambda(1-\lambda)}\,f(\lambda)$. The
period is then twice the time it takes to cross the chain,
\begin{equation}
T_\infty=2\int_0^{1}\frac{d\lambda}{2\sqrt{\lambda(1-\lambda)}\,f(\lambda)}
=\int_0^{\pi}\frac{d\theta}{f(\sin^2(\theta/2))},
\label{eq:Tinf}
\end{equation}
where the substitution $\lambda=(1-\cos\theta)/2$ removes the inverse
square-root endpoints. Evaluating Eq.~\eqref{eq:Tinf} on the limiting
profile gives $T_\infty=4.81193$, against the first revival times
$t_1=4.81052$, $4.81123$, and $4.81158$ computed at $L=512$, $1024$, and
$2048$, respectively. The difference halves at each doubling of $L$,
consistent with the $O(1/L)$ uniformity of the sampled spacings
(Appendix~\ref{app:ldos}).


\clearpage
\onecolumngrid

\setcounter{equation}{0}
\setcounter{figure}{0}
\renewcommand{\theequation}{S\arabic{equation}}
\renewcommand{\thefigure}{S\arabic{figure}}
\renewcommand{\theHequation}{S\arabic{equation}}
\renewcommand{\theHfigure}{S\arabic{figure}}

\begin{center}
\textbf{\large Supplemental Material for ``Exact solution of the forward-scattering approximation for the PXP model''}\\[2ex]
Saurav Pandey and Yevgeny Bar Lev\\[0.5ex]
\textit{Department of Physics, Ben-Gurion University of the Negev, Beer-Sheva 8410501, Israel}
\end{center}
\vspace{2ex}

This Supplemental Material collects the technical details behind the
hard-rod mapping of the main text, and also shows that the transfer of the
N\'eel state to its complement follows the same scaling as the revivals.
Section~\ref{sm:paths} gives closed forms for the zigzag fractions and the
rod weights, together with their large-$r$ behavior. Section~\ref{sm:rods}
explains why the particles of the gas are chosen to be the flipped sites
rather than the rods. Section~\ref{sm:transfer} compares the finite-$L$
recursion with direct construction of the FSA states and gives the norm at
the top of the FSA ladder in closed form. Section~\ref{sm:ends} derives the
two endpoints of the limiting hopping profile. Section~\ref{sm:pst} shows that the transfer peaks collapse onto the revival curve $\mathcal F(s)$.

\smsection{Closed forms for the zigzag fractions and the rod weights}
\label{sm:paths}

The fractions $p_m=\mathcal A_m/m!$ of the main text are the Taylor
coefficients of $\sec x+\tan x$ in Eq.~\eqref{eq:andre}. Note that the
generating function splits by parity: $\tan$ is odd and $\sec$ is even, so
the $\mathcal A_m$ with odd $m$ come from $\tan$ and count the orderings of
bulk domains, while those with even $m$ come from $\sec$ and count the
orderings of edge domains. The functions $\tan$ and $\sec$ are meromorphic
with poles at $x=(2k-1)\pi/2$ for $k\in\mathbb{Z}$, so each admits a
partial-fraction expansion over those poles~\cite{Flajolet2009}. Expanding
each pole term as a geometric series and collecting the coefficient of
$x^m$ turns the sum over $k$ into a Dirichlet series, giving closed-form
expressions for the bulk and edge fractions, respectively,
\begin{align}
p_{2r-1}&=2\Big(\tfrac{2}{\pi}\Big)^{2r}\big(1-2^{-2r}\big)\,\zeta(2r),
\label{eq:closedodd}\\
p_{2r}&=\tfrac{4}{\pi}\Big(\tfrac{2}{\pi}\Big)^{2r}\beta_D(2r+1),
\label{eq:closedeven}
\end{align}
with $\zeta$ the Riemann zeta function and $\beta_D$ the Dirichlet beta
function, $\beta_D(s)=\sum_{k\ge0}(-1)^k(2k+1)^{-s}$.

The rod weights of the main text, $w_r=p_{2r-1}^{2}$ and
$\tilde w_r=p_{2r}^{2}$, are therefore
\begin{align}
w_r&=4\Big(\tfrac{2}{\pi}\Big)^{4r}\big(1-2^{-2r}\big)^{2}\zeta(2r)^{2},
\label{eq:wbulk}\\
\tilde w_r&=\tfrac{16}{\pi^{2}}\Big(\tfrac{2}{\pi}\Big)^{4r}\beta_D(2r+1)^{2}.
\label{eq:wedge}
\end{align}
Since $\zeta(2r)$ and $\beta_D(2r+1)$ approach unity at large $r$, these
weights fall off geometrically as
\begin{equation}
w_r\to4v_c^{-r},
\qquad
\tilde w_r\to(4/\pi)^{2}v_c^{-r},
\qquad
v_c=(\pi/2)^{4},
\label{eq:wasymp}
\end{equation}
which is the behavior used in Appendix~\ref{app:eos} of the main text and
in Sec.~\ref{sm:ends}.

\smsection{The particle number of the gas}
\label{sm:rods}

In a hard-rod gas, the particle number is usually taken to be the number
of rods. In the main text, however, we take it to be the number of flipped
sites. The reason is that the FSA chain is indexed by $n$, and
Eq.~\eqref{eq:factorize} requires $Z_n$ at fixed $n$. The rod count is not
fixed by $n$. Moving one rung up the Hamming-distance ladder flips a
single site: if it is an $A$ site, it creates a new domain, and if it is a
$B$ site, whose two $A$ neighbors must already have flipped, it merges the
two domains ending at those neighbors into one. The number of rods thus
changes by $+1$ or $-1$, except at the right end, where flipping site $L$
turns a bulk domain into an edge domain and leaves the rod count
unchanged. The number of flipped sites, on the other hand, always changes
by exactly $+1$. Hence, in our convention, a rod may carry several
particles at once, $2r-1$ for a bulk rod of length $r$ and $2r$ for an
edge rod, as in Eq.~\eqref{eq:Zgas} of the main text.

\smsection{Check against direct construction, and the top of the FSA ladder}
\label{sm:transfer}

\begin{figure}[h]
\centering
\includegraphics{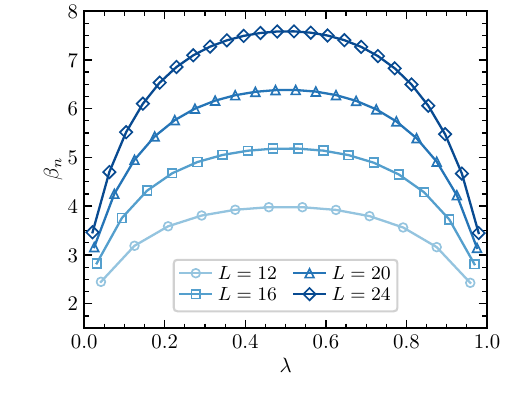}
\caption{Hopping coefficients $\beta_n$ of Eq.~\eqref{eq:betaZ} from the
hard-rod recursion of Eqs.~\eqref{eq:grec} and~\eqref{eq:transfer}
(lines), compared with those obtained from direct construction of
$(H^{+})^{n}\ket{v_0}$ in the constrained Hilbert space and
Eq.~\eqref{eq:fsachain} (open markers). Each $\beta_n$ is placed at the
bond midpoint $\lambda=(n+\tfrac12)/L$, and darker curves correspond to
larger $L$. The two routes give identical values at every $n$ and every
size shown.}
\label{fig:check}
\end{figure}

Figure~\ref{fig:check} tests the hard-rod mapping against direct
construction of the FSA states in the constrained Hilbert
space~\cite{Turner2018b}. The two agree exactly, with the norms $\Nn_n$
matching as integers, at every size tested. The recursion of
Eqs.~\eqref{eq:grec} and~\eqref{eq:transfer} of the main text, however,
costs only $O(L^3)$ operations and needs no Hilbert space at all.

One value, namely $\Nn_L$, follows from Eq.~\eqref{eq:transfer} without
any computation. At $n=L$, every site has flipped, so $S=\{1,\dots,L\}$
is a single domain of even length $L=2M$ terminating at site $L$. This is
an edge rod filling all $M$ cells, and there is exactly one such
configuration, so $Z_L=\tilde w_M=p_L^{2}$, which gives
\begin{equation}
\Nn_L=(L!)^{2}p_L^{2}=\mathcal A_L^{2}
\label{eq:NL}
\end{equation}
for the squared norm. The top of the FSA ladder is thus fixed by the
number of alternating permutations of $1,\dots,L$~\cite{Andre1879,Stanley2010},
which count the distinct paths from $\ket{v_0}$ to $\ket{v_L}$.

\smsection{The two endpoints of the limiting hopping profile}
\label{sm:ends}

Both ends of the limiting profile of Eq.~\eqref{eq:fprofile} can be
obtained by hand. Writing $W(x,u)=u^{-1}Q(xu^{2})$, as in
Eq.~\eqref{eq:Qdef}, turns the first relation of the equation of state,
$x+W(x,u)=1$ in Eq.~\eqref{eq:eos}, into
\begin{equation}
v+u\,Q(v)=u^{2},
\label{eq:eosv}
\end{equation}
an equation in $v$ alone, whose root $v_*$ gives $x_*=v_*/u^{2}$, as in
Appendix~\ref{app:eos}. The density of Eq.~\eqref{eq:eos} then becomes
\begin{equation}
\rho(u)=-\frac{\partial\ln x_*}{\partial\ln u}
=2-\frac{\partial\ln v_*}{\partial\ln u},
\label{eq:rhov}
\end{equation}
so that it suffices to find $v_*(u)$. Setting $\rho(u_*)=2\lambda$ then
fixes $u_*(\lambda)$, which, through Eq.~\eqref{eq:fprofile}, gives $f(\lambda)$ explicitly near $\lambda=0$ in the dilute limit and near $\lambda=1$ in the close-packed limit.

\subsection*{The dilute end}

As $\lambda\to0$, the gas is nearly empty, and thus the fugacity $u$ is
small. Since the root of Eq.~\eqref{eq:eosv} is $v_*=O(u^{2})$, the
$r$-th term of $u\,Q(v_*)$ is $O(u^{2r+1})$, so rods with $r\ge2$
contribute only at higher order. Keeping only the $r=1$ term, with
$w_1=1$ from Eq.~\eqref{eq:wbulk}, Eq.~\eqref{eq:eosv} becomes linear,
$v_*+uv_*=u^{2}$, with root $v_*=u^{2}/(1+u)$. Equation~\eqref{eq:rhov}
then yields the density $\rho=u/(1+u)$, and setting $\rho=2\lambda$ fixes
$u_*=2\lambda/(1-2\lambda)$. Equation~\eqref{eq:fprofile} finally gives
\begin{equation}
f^{2}_{\rm dilute}(\lambda)=\frac{1-2\lambda}{2(1-\lambda)}
=\tfrac12\left(1-\lambda\right)+O(\lambda^{2}),
\label{eq:diluteprofile}
\end{equation}
so that $f(0)=1/\sqrt{2}$.

\subsection*{The close-packed end}

As $\lambda\to1$, the fugacity grows without bound, and the root $v_*$
approaches $v_c$, where $Q$ diverges. With
$(1-2^{-2r})\,\zeta(2r)=1+3^{-2r}+5^{-2r}+\dots$, Eq.~\eqref{eq:wbulk}
becomes $w_r=4v_c^{-r}\big[1+2\cdot9^{-r}+\dots\big]$. The leading term
sums to $4\sum_r(v/v_c)^{r}=4v/(v_c-v)$, which diverges at $v_c$, whereas
the corrections sum to a function that remains finite there. Near $v_c$,
therefore, $Q(v)=4v/(v_c-v)+O(1)$. Writing $v_*=v_c(1-\delta)$, we have
$Q(v_*)\simeq4/\delta$, and at large $u$, Eq.~\eqref{eq:eosv} reduces to
$4u/\delta\simeq u^{2}$, so that $\delta\simeq4/u$. Then $\ln v_*=\ln v_c+\ln(1-\delta)\simeq\ln v_c-4/u$, and Eq.~\eqref{eq:rhov} yields the density $\rho\simeq2-4/u$. Setting $\rho=2\lambda$ fixes $u_*\simeq2/(1-\lambda)$,
and Eq.~\eqref{eq:fprofile} finally gives
\begin{equation}
f^{2}_{\rm dense}(\lambda)\simeq\frac{\lambda}{2},
\label{eq:denseprofile}
\end{equation}
so that $f(1)=1/\sqrt{2}$ as well.

\smsection{Transfer of the N\'eel state to its complement}
\label{sm:pst}

\begin{figure}[h]
  \centering
  \includegraphics{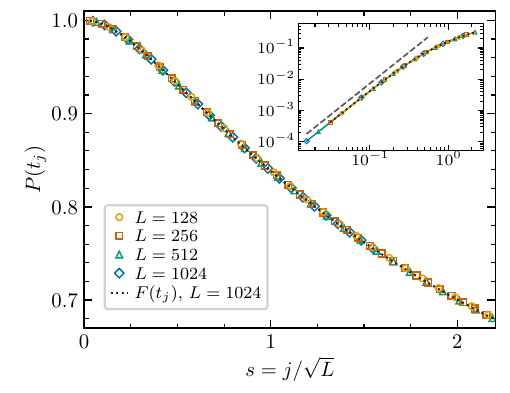}
  \caption{Transfer of the N\'eel state to $\ket{v_L}$ within the FSA. Main panel: the
  peak values $P(t_j)$ against $s=j/\sqrt L$, with half-integer $j$. The dotted line is
  the revival curve $F(t_j)$ of Fig.~\ref{fig:main}(c) at $L=1024$. Inset: $1-P(t_j)$
  on logarithmic axes, with the dashed line $\propto s^2$.}
  \label{fig:transfer}
\end{figure}

Within the FSA, the N\'eel state $\ket{v_0}$ is transferred to its
complement $\ket{v_L}$, at the other end of the chain, halfway between consecutive revivals. The transfer fidelity
\begin{equation}
P(t)=\big|\bra{v_L}\ee^{-iH_{\rm FSA}t}\ket{v_0}\big|^{2}
\label{eq:Pt}
\end{equation}
peaks at times $t_j\simeq jT$, with $T$ the revival period and half-integer
$j=\tfrac12,\tfrac32,\dots$, so that $j$ counts the elapsed time in revival
periods, as for the revivals in Fig.~\ref{fig:main}(c) of the main text.
Figure~\ref{fig:transfer} shows that, with this labeling, the peak values
$P(t_j)$ collapse onto the revival curve $\mathcal F(s)$, with
$s=j/\sqrt L$ and $1-P(t_j)\propto s^{2}$ at small $s$. At fixed $j$, the
deficit of each transfer is therefore $O(j^{2}/L)$, and, like the revivals,
the transfers become perfect as $L\to\infty$.

\end{document}